\documentclass[fleqn,usenatbib]{mnras}

\usepackage[T1]{fontenc}
\usepackage{newtxtext,newtxmath}
\usepackage{graphicx}
\usepackage{amsmath}
\usepackage{bm}
\usepackage{xcolor}

\newcommand{\figalttext}[1]{\par\smallskip\noindent\textit{Alt text: #1}}

\begin{document}
\label{firstpage}
\pagerange{\pageref{firstpage}--\pageref{lastpage}}

\title[Tests of smooth sign-switching $\Lambda_s$CDM]{Catching the Cosmic Sign Flip: Background and Growth Tests of Smooth Sign Switching \texorpdfstring{$\Lambda_s$}{Lambda\_s}CDM}

\author[A. Mitra]{Ayan Mitra$^{1,2}$\thanks{E-mail: \href{mailto:ayan@illinois.edu}{ayan@illinois.edu}}\\
$^{1}$Center for AstroPhysical Surveys, National Center for Supercomputing Applications, Urbana, IL 61801, USA\\
$^{2}$Department of Astronomy, University of Illinois at Urbana-Champaign, Urbana, IL 61801, USA}

\date{Accepted XXX. Received YYY; in original form ZZZ}
\pubyear{\the\year{}}

\maketitle

\begin{abstract}
Recent baryon acoustic oscillation (BAO) measurements from the Dark Energy Spectroscopic Instrument (DESI), in combination with cosmic microwave background (CMB) and supernova data, have strengthened the case for testing dynamical dark energy extensions to flat $\Lambda$CDM. In analyses that allow an effective neutrino mass parameter to cross zero, these data can also drive the best fit towards an unphysical negative value, motivating models that mimic this effect through late time expansion physics. The sign switching cosmological constant ($\Lambda_s$CDM) model is one such phenomenological framework, introducing an AdS like negative vacuum energy density at higher redshifts ($z > z^\dagger$) that transitions to a dS like positive vacuum energy density at low redshifts. Many phenomenological fits of the original $\Lambda_s$CDM model assume a discontinuous step function transition. In this work, we generalise the transition dynamics using a smooth parametrisation, $\Lambda(z) \propto \tanh[\Delta(z^\dagger - z)]$, and constrain the transition redshift $z^\dagger$ and smoothness parameter $\Delta$ using joint background distance data from DES-SN5YR with the Dovekie recalibration of DES and DESI 2024 BAO. We further incorporate a compact redshift space distortion (RSD) $f\sigma_8$ compilation and show that current growth data do not break the transition sharpness degeneracy. The joint data do not prefer the smooth sign switching model over flat $\Lambda$CDM: the $\chi^2_{\text{min}}$ values are nearly identical, while information criteria penalise the additional parameters. The transition smoothness parameter remains dominated by prior volume and unconstrained ($\Delta = 24.75^{+17.42}_{-17.48}$ for the joint fit), indicating that current late time data cannot distinguish between a sharp phase transition and a smooth dynamical crossover. This work does not directly constrain the physical sum of neutrino masses $\sum m_\nu$; rather, it tests whether late time background and growth data alone can constrain the transition dynamics. We discuss the physical implications of this null detection and present projections for next generation surveys.
\end{abstract}


\section{Introduction}
\label{sec:intro}
The flat $\Lambda$CDM model has served as the standard model of cosmology for nearly three decades, successfully explaining a wide range of observations from the cosmic microwave background (CMB; e.g., \citep{Planck:2018vyg, Benson:2014qxb}) to the large scale structure of the Universe. However, the increase in observational precision has revealed persistent tensions between early and late time probes. The most notable are the $H_0$ tension, a $5\sigma$ discrepancy between CMB inferred values and local measurements from Cepheid calibrated distance ladders \citep{Riess:2021jrx} and Tip of the Red Giant Branch (TRGB) calibrations \citep{Freedman:2019jwv, DiValentino:2021izs}, and the $S_8$ tension in the growth of structure \citep{DES:2021wwg}. Early Universe solutions, such as early dark energy (EDE) \citep{Poulin:2018cxd, Karwal:2016vyg}, seek to address this by accelerating expansion prior to recombination.

The first year DESI BAO measurements introduced a related challenge: when combined with CMB and Type Ia supernova data, models with varying dark energy show a mild to moderate preference over flat $\Lambda$CDM, depending on the supernova compilation used \citep{DESI:2024mwt}. Recent analyses using DESI DR2, CMB data and the Dovekie recalibration of DES supernovae have also highlighted a neutrino mass boundary tension: when an effective neutrino mass parameter is allowed to continue below zero, the flat $\Lambda$CDM best fit can move to $\sum m_{\nu,\text{eff}} < 0$ \citep{Kibris:2026neg}. This negative effective mass should not be interpreted as a physical neutrino mass, but as a diagnostic of missing expansion or growth physics.

A phenomenological dark energy model proposed to address this boundary tension is the sign switching cosmological constant model ($\Lambda_s$CDM) \citep{Akarsu:2021txy,Akarsu:2022lhx,Akarsu:2023xgf,Kibris:2026neg}. By replacing the standard constant vacuum energy with an effective cosmological constant that switches sign from negative (anti de Sitter, AdS like phase) in the past to positive (de Sitter, dS like phase) at a transition redshift $z^\dagger \sim 2$, the model alters the expansion rate at intermediate redshifts. Full CMB + BAO + supernova fits can then move the effective neutrino mass constraint back into the physical positive region \citep{Kibris:2026neg}.

\citet{Bouhmadi-Lopez:2025cos} recently performed a comprehensive perturbative study of several sign switching dark energy scenarios, exploring the linear perturbation equations and growth observables such as the matter contrast ($\delta_m$), gravitational potential ($\Phi$), matter power spectrum ($P(k)$), and growth rate ($f\sigma_8$). In this work, we complement their physical analysis of growth observables with joint cosmological parameter estimation. We incorporate both background distance data (Dovekie recalibrated DES supernovae and DESI 2024 BAO) and linear growth data (RSD $f\sigma_8$) to constrain the transition parameters $(z^\dagger, \Delta)$ alongside the growth parameter $\sigma_8$. Rather than solving the full relativistic perturbation equations across all scales, we implement an efficient vectorised subhorizon linear growth solver. This allows us to quantify how much information current background and compressed growth data carry about the smooth sign switching transition. We summarise the differences in scope and methodology in Table \ref{tab:lit_comparison}.

\begin{table*}[t]
\centering
\caption{\label{tab:lit_comparison} Comparison of this work with the recent perturbative analysis of smooth sign switching cosmologies by \protect\citet{Bouhmadi-Lopez:2025cos}.}
\begin{tabular}{lccccc}
Work & Models & Main Observables & Perturbations/Growth? & Data Use & Main Role \\
\hline
\parbox[t]{3.2cm}{\citet{Bouhmadi-Lopez:2025cos}} & 
\parbox[t]{2.2cm}{Abrupt and smooth sign switching models} & 
$\delta_m$, $\Phi$, $P(k)$, $f\sigma_8$ & 
\parbox[t]{3cm}{Full linear perturbation equations} & 
\parbox[t]{3.2cm}{Growth/observational comparison} & 
\parbox[t]{3cm}{Perturbative viability} \\
\hline
This work & 
\parbox[t]{2.2cm}{Smooth tanh $\Lambda_s$C{D}M with $(z^\dagger, \Delta)$} & 
\parbox[t]{2.5cm}{SN, BAO, and RSD distances and $f\sigma_8$} & 
\parbox[t]{3cm}{Linear subhorizon growth\footnotemark[1]} & 
\parbox[t]{3.2cm}{DES-SN5YR + DESI BAO + RSD compilation} & 
\parbox[t]{3cm}{Joint background + growth constraints} \\
\end{tabular}
\footnotetext[1]{We solve the subhorizon linear growth ODE using a vectorised RK4 algorithm and fit RSD $f\sigma_8$ measurements.}
\end{table*}

Many phenomenological fits of the original $\Lambda_s$CDM model have treated the sign switch as a discontinuous step function. Here, we focus on a tanh smoothing of that transition and ask whether current data can constrain its width. We first establish baseline background only constraints using the DES-SN5YR sample with the Dovekie recalibration of DES and the DESI 2024 BAO measurements. We then add a compact RSD $f\sigma_8$ likelihood using a subhorizon growth solver to test whether current growth data improve the transition constraints. This approach allows us to place joint background + growth MCMC constraints on the transition redshift $z^\dagger$ and transition smoothness parameter $\Delta$, evaluating whether current observations show any preference for this transition or can constrain its sharpness.

\section{Theoretical Framework}
\label{sec:theory}
We consider a flat FLRW universe where the Friedmann equation is written as:
\begin{equation}
H^2(z) = H_0^2 \left[ \Omega_{m0}(1+z)^3 + \Omega_{r0}(1+z)^4 + \Omega_{\Lambda}(z) \right],
\end{equation}
where $\Omega_{m0}$ and $\Omega_{r0}$ represent the present day matter and radiation density parameters, and $\Omega_{\Lambda}(z) = \rho_{\Lambda}(z)/\rho_{\text{crit},0}$ is the effective dark energy density parameter.

To model a smooth mirror transition from AdS to dS, we adopt a hyperbolic tangent profile (a standard sigmoid smoothing function used in transitional dark energy literature, see e.g., \citep{Akarsu:2021txy, Akarsu:2023xgf, Bouhmadi-Lopez:2025cos}):
\begin{equation}
\Omega_{\Lambda}(z) = (1 - \Omega_{m0} - \Omega_{r0}) \frac{\tanh[\Delta(z^\dagger - z)]}{\tanh[\Delta z^\dagger]},
\label{eq:omega_lambda}
\end{equation}
where $z^\dagger$ is the transition redshift where the vacuum energy density crosses zero, and $\Delta > 0$ controls the rate (sharpness) of the transition. The normalisation factor $\tanh[\Delta z^\dagger]$ ensures that the model exactly satisfies the flat Universe constraint $\Omega_{\Lambda}(0) = 1 - \Omega_{m0} - \Omega_{r0}$ at the present day ($z=0$).
In the limit $\Delta \to \infty$, this smooth profile converges to the abrupt signum like transition model:
\begin{equation}
\Omega_{\Lambda}(z) \to (1 - \Omega_{m0} - \Omega_{r0}) \operatorname{sgn}(z^\dagger - z).
\end{equation}

If the vacuum energy density depends on redshift, $\rho_\Lambda(z) = \rho_{\text{crit},0} \Omega_\Lambda(z)$, it must satisfy the covariant conservation equation of the total stress energy tensor, $T^{\mu\nu}_{;\nu} = 0$. If we treat dark energy as an isolated effective fluid that is individually conserved, its energy density must satisfy the continuity equation $\dot{\rho}_{\text{DE}} + 3H(\rho_{\text{DE}} + p_{\text{DE}}) = 0$. This implies an effective equation of state $w_{\text{eff}}(z) \equiv p_{\text{DE}}/\rho_{\text{DE}}$ given by:
\begin{equation}
w_{\text{eff}}(z) = -1 + \frac{1}{3}\frac{d\ln\rho_{\text{DE}}}{d\ln(1+z)}.
\end{equation}
For our smooth hyperbolic tangent parametrisation, this yields:
\begin{equation}
w_{\text{eff}}(z) = -1 + \frac{\Delta (1+z) \operatorname{sech}^2[\Delta(z^\dagger - z)]}{3 \tanh[\Delta(z^\dagger - z)]}.
\label{eq:w_eff}
\end{equation}
In the limits far from the transition ($z \ll z^\dagger$ and $z \gg z^\dagger$), the effective equation of state approaches $w_{\text{eff}}(z) \to -1$, behaving as a positive (dS like) or negative (AdS like) cosmological constant, respectively. At the transition point $z = z^\dagger$, the effective equation of state diverges because the density $\rho_{\text{DE}}(z^\dagger) = 0$ crosses zero, which is a common mathematical artefact of zero crossing models. However, the physical pressure remains smooth and finite everywhere:
\begin{equation}
p_{\text{eff}}(z) = -\rho_{\Lambda 0} \left[ \frac{\tanh[\Delta(z^\dagger - z)]}{\tanh[\Delta z^\dagger]} + \frac{\Delta (1+z) \operatorname{sech}^2[\Delta(z^\dagger - z)]}{3 \tanh[\Delta z^\dagger]} \right],
\end{equation}
where $\rho_{\Lambda 0} \equiv \rho_{\text{crit},0}(1 - \Omega_{m0} - \Omega_{r0})$.

Alternatively, if dark energy is interpreted as a varying vacuum energy $\Lambda(z)$ directly coupled to the matter sector, the total stress energy tensor is conserved through energy exchange, satisfying:
\begin{equation}
\dot{\rho}_m + 3H\rho_m = -\dot{\rho}_\Lambda.
\end{equation}
This implies continuous matter creation or annihilation as the vacuum energy evolves, which has been studied in decaying vacuum cosmologies (e.g., $\Lambda(t)\text{CDM}$ \citep{Planck:2018vyg}). Throughout this work, we treat $\Omega_{\Lambda}(z)$ as an effective dark energy component at the background level. The smooth transition profile with finite $\Delta$ is physically motivated by the need to avoid the step discontinuity of the abrupt model ($\Delta \to \infty$), which has been argued to introduce singular behaviour in curvature perturbations and cosmic growth equations near the jump (see e.g., \citep{Akarsu:2021txy,Akarsu:2023xgf}). Moreover, a fully consistent perturbation treatment of a fluid crossing density zero is challenging; the divergence of $w_{\text{eff}}$ at $\rho_{\text{DE}}=0$ can make perturbations mathematically subtle or poorly defined. While setting the effective sound speed to $c_s^2 = 1$ is a standard choice to avoid clustering instabilities, it is not a complete resolution. A complete perturbation framework for sign switching cosmologies near the zero crossing requires further study. For the compact growth analysis performed here, we therefore use a vectorised subhorizon growth solver as a controlled phenomenological test rather than a replacement for a full Einstein Boltzmann treatment.

\section{Data and Methodology}
\label{sec:data}
We constrain the cosmological parameters using two distinct setups: a background only analysis with parameter space $\boldsymbol{\theta}_{\text{bg}} = \{\Omega_{m0}, \mathcal{K}, z^\dagger, \Delta\}$, and a joint background + growth analysis with parameter space $\boldsymbol{\theta}_{\text{joint}} = \{\Omega_{m0}, \mathcal{K}, z^\dagger, \Delta, \sigma_8\}$. The calibration factor is defined as $\mathcal{K} \equiv c / (H_0 r_d)$. The data sets used are:
\begin{itemize}
\item \textbf{Type Ia Supernovae:} We use the DES-SN5YR sample with the Dovekie recalibration of DES, comprising roughly 1,600 likely DES SNe Ia and roughly 200 low redshift SNe Ia over $0.02 < z < 1.13$ \citep{Popovic:2025dovekie}. This builds upon previous samples such as Pantheon+ \citep{Brout:2022vvo}. Supernova light curves are fitted using the SALT2 \citep{Guy:2007zz} and SALT3 \citep{Kenworthy:2021zwo} templates. We use the full covariance matrix incorporating both statistical and systematic uncertainties, and analytically marginalise over the absolute magnitude $M$ to eliminate dependence on the Hubble parameter $H_0$.
\item \textbf{Baryon Acoustic Oscillations:} We use the DESI 2024 BAO data vector consisting of 12 measurements of spherically averaged volume distances ($D_V/r_d$), transverse comoving distances ($D_M/r_d$), and Hubble distances ($D_H/r_d$) extracted from the BGS, LRG, ELG, QSO, and Lyman-$\alpha$ tracers spanning the redshift range $0.1 < z < 4.2$ \citep{DESI:2024mwt}, along with their joint covariance matrix.
\item \textbf{Redshift Space Distortions (RSD):} To constrain the growth of cosmic structure, we incorporate a compilation of 7 effective growth rate measurements $f\sigma_8(z)$ covering $0.15 \le z \le 2.33$ derived from the SDSS MGS \citep{Howlett:2015aoa}, BOSS DR12 \citep{BOSS:2016wmc}, and the final eBOSS \citep{eBOSS:2020yzd} galaxy and quasar surveys. The data vector consists of effective redshifts $z_{\text{eff}} = \{0.15, 0.38, 0.51, 0.698, 0.85, 1.48, 2.33\}$ and measured values $f\sigma_{8,\text{obs}} = \{0.530 \pm 0.160, 0.440 \pm 0.040, 0.458 \pm 0.038, 0.473 \pm 0.033, 0.315 \pm 0.095, 0.462 \pm 0.045, 0.387 \pm 0.081\}$. We treat the seven RSD measurements as independent effective constraints; a future full shape analysis should use the survey covariance consistently.
\end{itemize}

To evaluate the growth observables in our MCMC pipeline, we solve the subhorizon linear growth factor $D(a)$ equation:
\begin{equation}
\frac{d^2D}{dx^2} + \left(2 + \frac{d\ln H}{dx}\right) \frac{dD}{dx} - \frac{3}{2}\Omega_m(x) D = 0,
\label{eq:growth_ode}
\end{equation}
where $x \equiv -\ln(1+z) = \ln a$, and the matter density fraction is $\Omega_m(x) = \Omega_{m0} e^{-3x}/E^2(x)$. We solve this second order ODE using an optimised vectorised fourth order Runge Kutta (RK4) solver integrated from $x = -4$ ($z \approx 53.6$, deep in the matter dominated epoch where $D(a) \propto a$) to the present day $x = 0$ ($z = 0$). The initial conditions at $x_{\text{ini}} = -4.0$ are set to $D(x_{\text{ini}}) = e^{x_{\text{ini}}}$ and $dD/dx(x_{\text{ini}}) = e^{x_{\text{ini}}}$, which is consistent with the standard matter dominated growing mode attractor solution $D(a) \propto a$. The growth rate is defined as $f(z) \equiv d\ln D/d\ln a$, which allows us to compute the growth observable:
\begin{equation}
f\sigma_8(z) = \sigma_8 \cdot f(z) \cdot \frac{D(z)}{D(0)}.
\end{equation}

We sample the posterior distributions using the affine invariant Markov Chain Monte Carlo (MCMC) sampler \texttt{emcee} \citep{Foreman-Mackey:2012oct} (cross validated with the \texttt{cobaya} package \citep{Torrado:2020dgo} and analysed using \texttt{GetDist} \citep{Lewis:2019xzd}) with flat priors:
\begin{eqnarray}
\Omega_{m0} &\in& [0.05, 0.95], \\
\mathcal{K} &\in& [10.0, 50.0], \\
z^\dagger &\in& [0.1, 5.0], \\
\Delta &\in& [0.1, 50.0], \\
\sigma_8 &\in& [0.4, 1.2].
\end{eqnarray}

Fitting $\mathcal{K} \equiv c / (H_0 r_d)$ as a free, unconstrained calibration parameter is equivalent to treating the sound horizon at the drag epoch $r_d$ as completely free. This compressed BAO approach is intentionally independent of early Universe physics, CMB measurements, and Big Bang Nucleosynthesis (BBN) constraints. By doing so, our results isolate late time geometric constraints on the expansion history $H(z)$ and transition dynamics $(z^\dagger, \Delta)$. If we were instead to impose a tight CMB prior on $r_d$ (e.g., $r_d = 147.09 \pm 0.26$ Mpc from Planck \citep{Planck:2018vyg}), it would introduce a strong correlation between $H_0$ and the sound horizon. In models that reduce $H(z)$ at early times (such as $\Lambda_s$CDM at $z > z^\dagger$), a fixed CMB calibration would shift $H_0$ to higher values or require a physically consistent, positive sum of neutrino masses ($\sum m_\nu > 0$) to maintain consistency with the measured acoustic scale. By keeping $\mathcal{K}$ free, we ensure that our background level constraints are conservative and represent the minimum leverage of the distance data alone.

For our primary analyses, we run the MCMC sampler with 32 walkers for 6,000 steps per walker (and 30,000 steps for final convergence runs), discarding the first 1,000 steps as burn in. The convergence is monitored by calculating the integrated autocorrelation time $\tau$ for each parameter. The maximum autocorrelation time obtained is $\tau_{\text{max}} \approx 88.6$ steps (associated with the transition smoothness parameter $\Delta$). The ratio of the chain length after burn in per walker to the maximum autocorrelation time is well above the standard heuristic threshold of 50. The average acceptance fraction across all walkers is $28.4\%$, within the typical target range of $20\%$ to $50\%$ for the affine invariant algorithm. Visual inspection of the trace plots and Gelman Rubin statistics ($R-1 < 0.01$ across chains) indicate well mixed, unimodal posteriors.

To perform a Bayesian model comparison, we also compute the Bayesian evidence (or marginal likelihood) $\log Z$:
\begin{equation}
Z = \int \mathcal{L}(\mathbf{d}|\boldsymbol{\theta}) \pi(\boldsymbol{\theta}) d\boldsymbol{\theta},
\end{equation}
where $\mathcal{L}$ is the likelihood and $\pi$ is the prior distribution. We solve this multidimensional integral using the nested sampling package \texttt{dynesty} \citep{Speagle:2020dyn} with a multi ellipsoidal slice sampling algorithm and $N_{\text{live}} = 250$ live points, which provides the evidence $\log Z$ alongside its numerical uncertainty.

\section{Results}
\label{sec:results}
Before presenting the full MCMC results, it is useful to establish a simple analytic argument for why late time background distance data are expected to be insensitive to the transition sharpness $\Delta$ when the transition redshift occurs in the matter dominated epoch ($z^\dagger \gtrsim 2$). Under flat $\Lambda_s$CDM, the effective dark energy density fraction is given by Eq.~(\ref{eq:omega_lambda}). In the matter dominated regime ($z \gg 1$), the ratio of dark energy to matter density scales as:
\begin{equation}
\begin{aligned}
\frac{\Omega_\Lambda(z)}{\Omega_m(z)}
&\approx \frac{1 - \Omega_{m0}}{\Omega_{m0}}
\frac{\operatorname{sgn}(z^\dagger - z)}{(1+z)^3} \\
&\approx \frac{0.7}{0.3}\frac{1}{(1+z)^3}
\approx \frac{2.3}{(1+z)^3}.
\end{aligned}
\end{equation}
At $z \approx 3.5$, this ratio is $\Omega_\Lambda(3.5)/\Omega_m(3.5) \approx 2.3 / 91 \approx 0.025$, meaning that dark energy represents less than $3\%$ of the total energy density of the Universe. The comoving angular diameter distance $D_M(z)$ integrates over $1/H(z')$, so the impact of the transition details (such as its width $\delta z \sim 1/\Delta$) on the distance observables is suppressed by both the integral comoving distance projection and matter dominance at high redshifts. Since the maximum redshift of the DES-SN5YR sample is $z \approx 1.13$ and the DESI 2024 BAO measurements at high redshifts have statistical errors of several percent, any localised changes in the expansion history around $z \approx 3.5$ are expected to be below the observational noise level. This insensitivity implies that MCMC fits should naturally yield posteriors on $\Delta$ that are dominated by the prior; the numerical result is therefore a confirmation of this analytic expectation rather than a surprising detection.

\begin{figure*}
\centering
\includegraphics[width=0.49\textwidth]{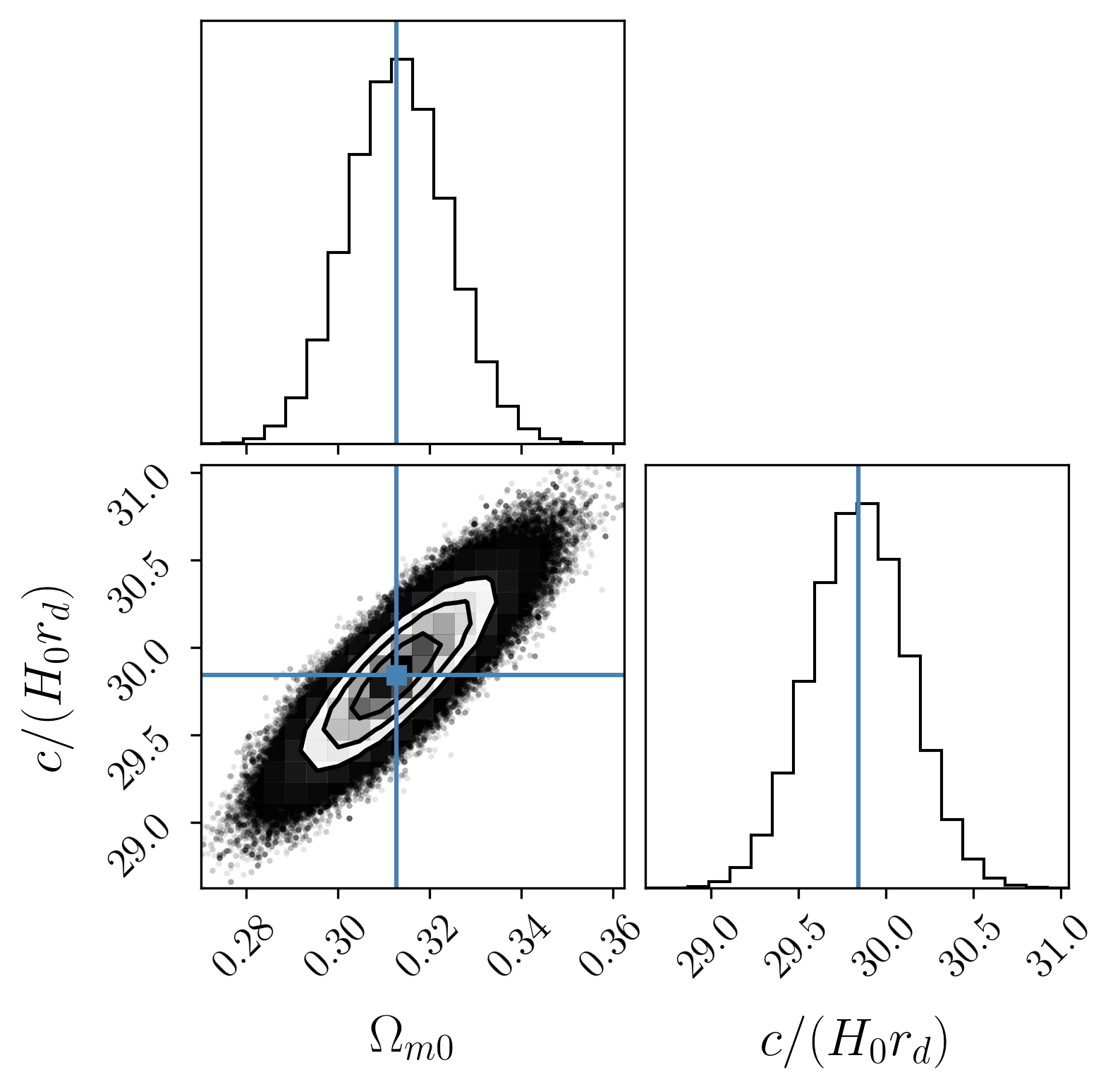}
\hfill
\includegraphics[width=0.49\textwidth]{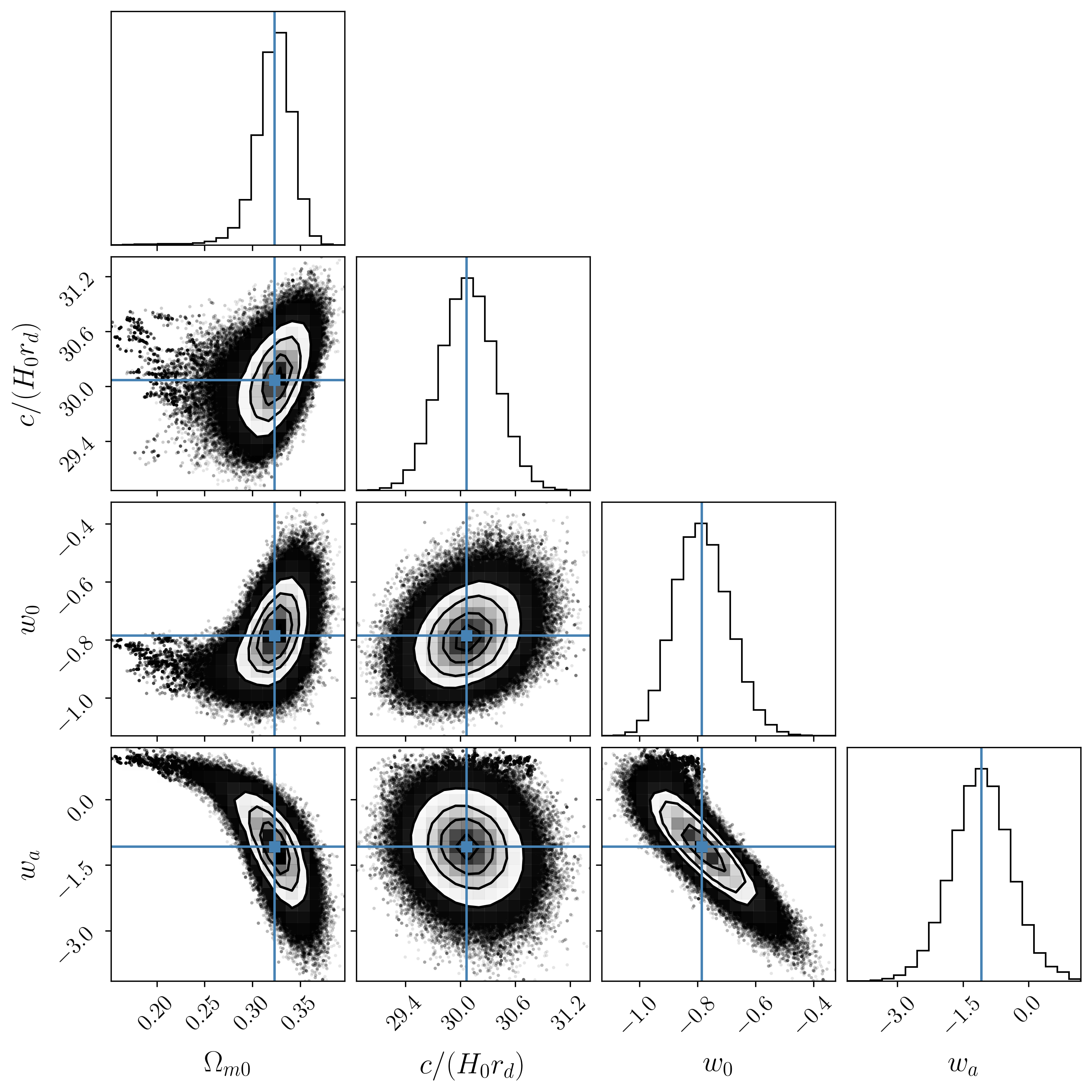}
\caption{Joint posterior distributions (corner plots) \protect\citep{Foreman-Mackey:2016corner} for flat $\Lambda$CDM (left) and CPL (right). The contours show $68\%$ and $95\%$ confidence regions, and the black markers represent the maximum likelihood points.}
\figalttext{Two corner plots comparing posterior constraints for flat $\Lambda$CDM and CPL. Each panel shows one- and two-dimensional marginalized distributions with confidence contours and maximum-likelihood markers.}
\label{fig:corner_lcdm_cpl}
\end{figure*}

\begin{figure*}
\centering
\includegraphics[width=0.66\textwidth]{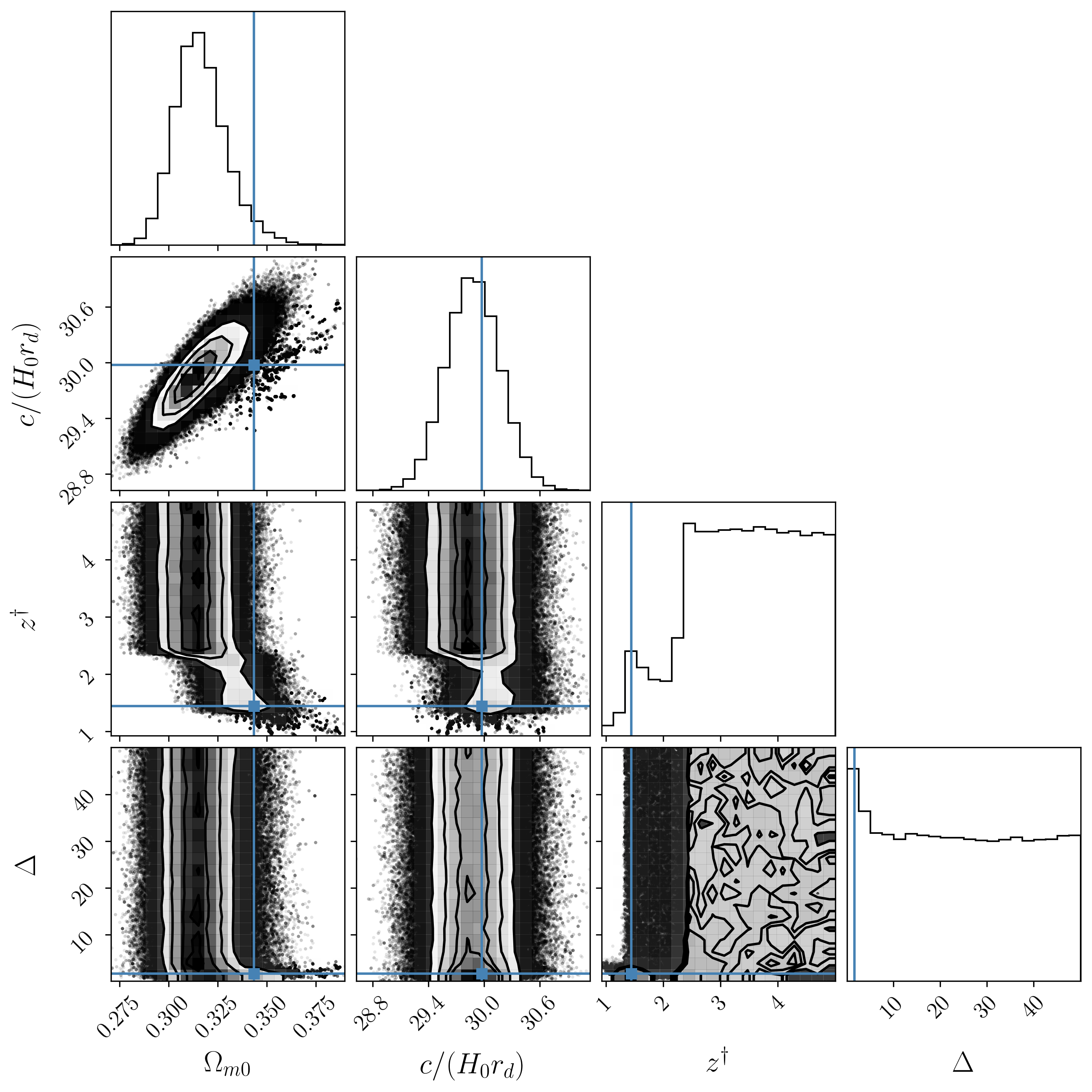}
\caption{Joint posterior distribution (corner plot) \protect\citep{Foreman-Mackey:2016corner} for the smooth sign switching $\Lambda_s$CDM model. The contours show $68\%$ and $95\%$ confidence regions for $\Omega_{m0}$, $\mathcal{K} = c/(H_0 r_d)$, $z^\dagger$, and $\Delta$; the black markers represent the maximum likelihood point.}
\figalttext{Corner plot of posterior constraints for the smooth sign switching $\Lambda_s$CDM model, showing marginalized distributions and confidence contours for the matter density, BAO calibration, transition redshift, and transition smoothness.}
\label{fig:corner_lambda_s}
\end{figure*}

\begin{figure*}
\centering
\includegraphics[width=0.49\textwidth]{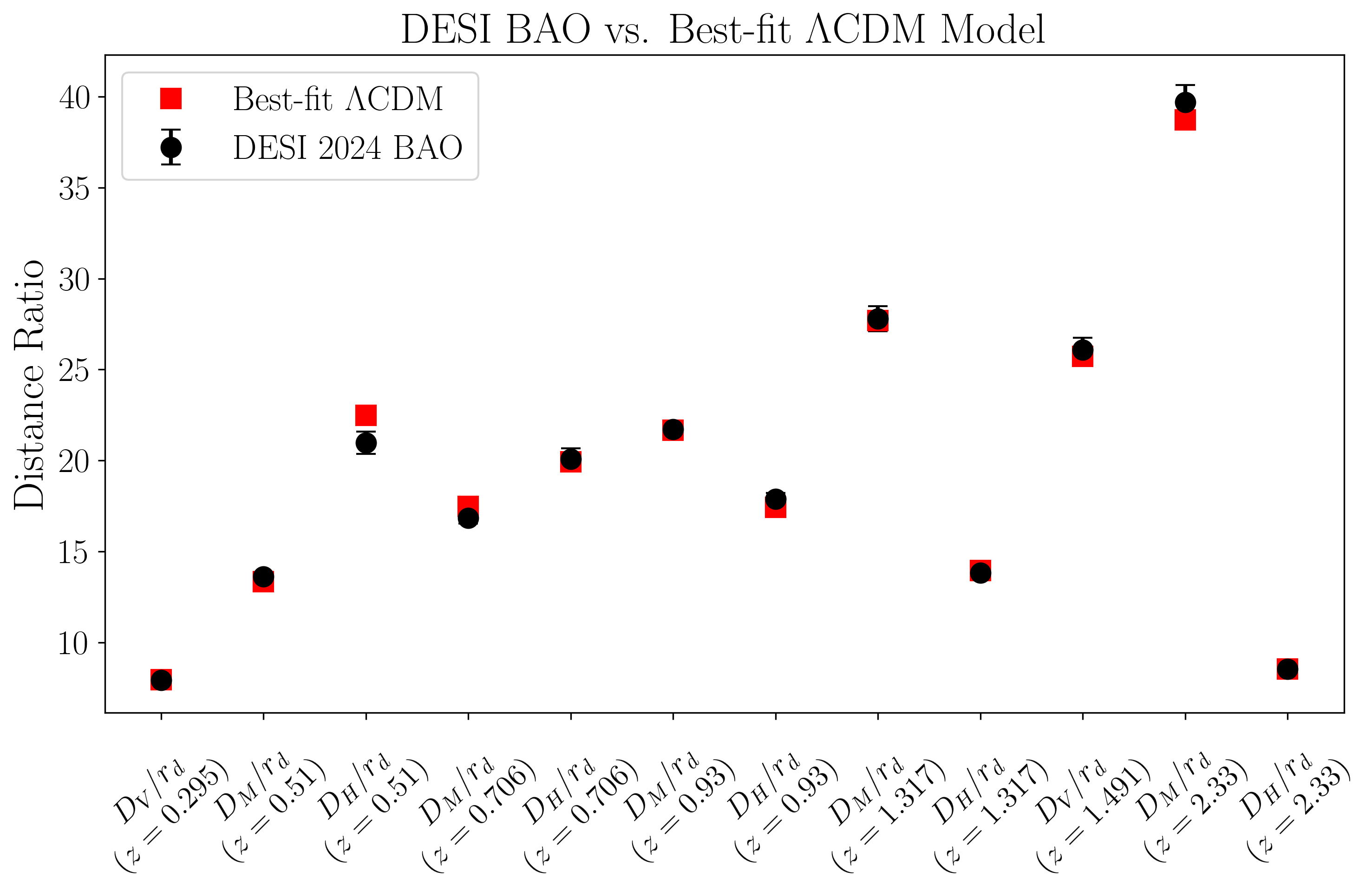}
\hfill
\includegraphics[width=0.49\textwidth]{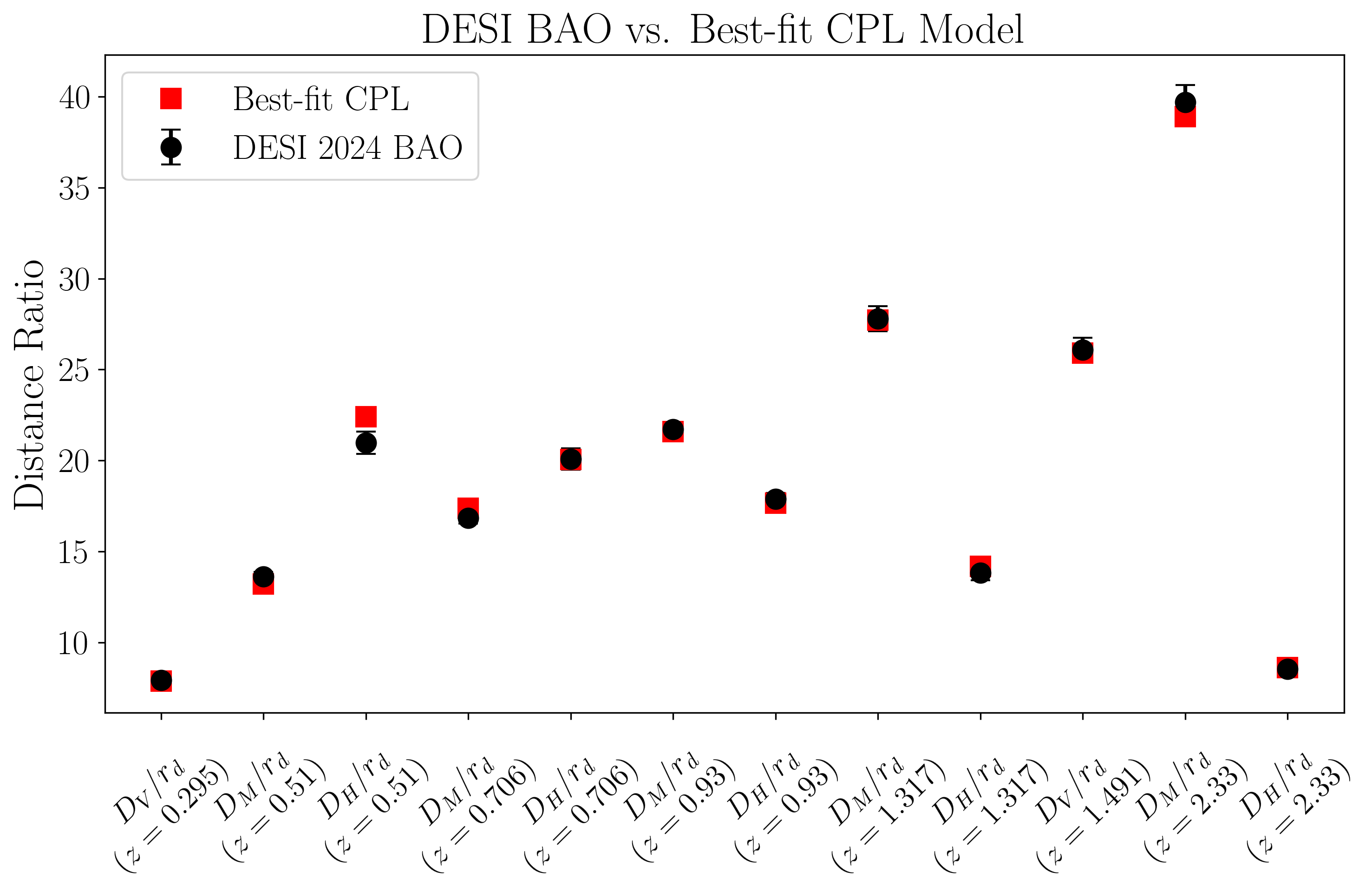}
\\ \vspace{0.4cm}
\includegraphics[width=0.7\textwidth]{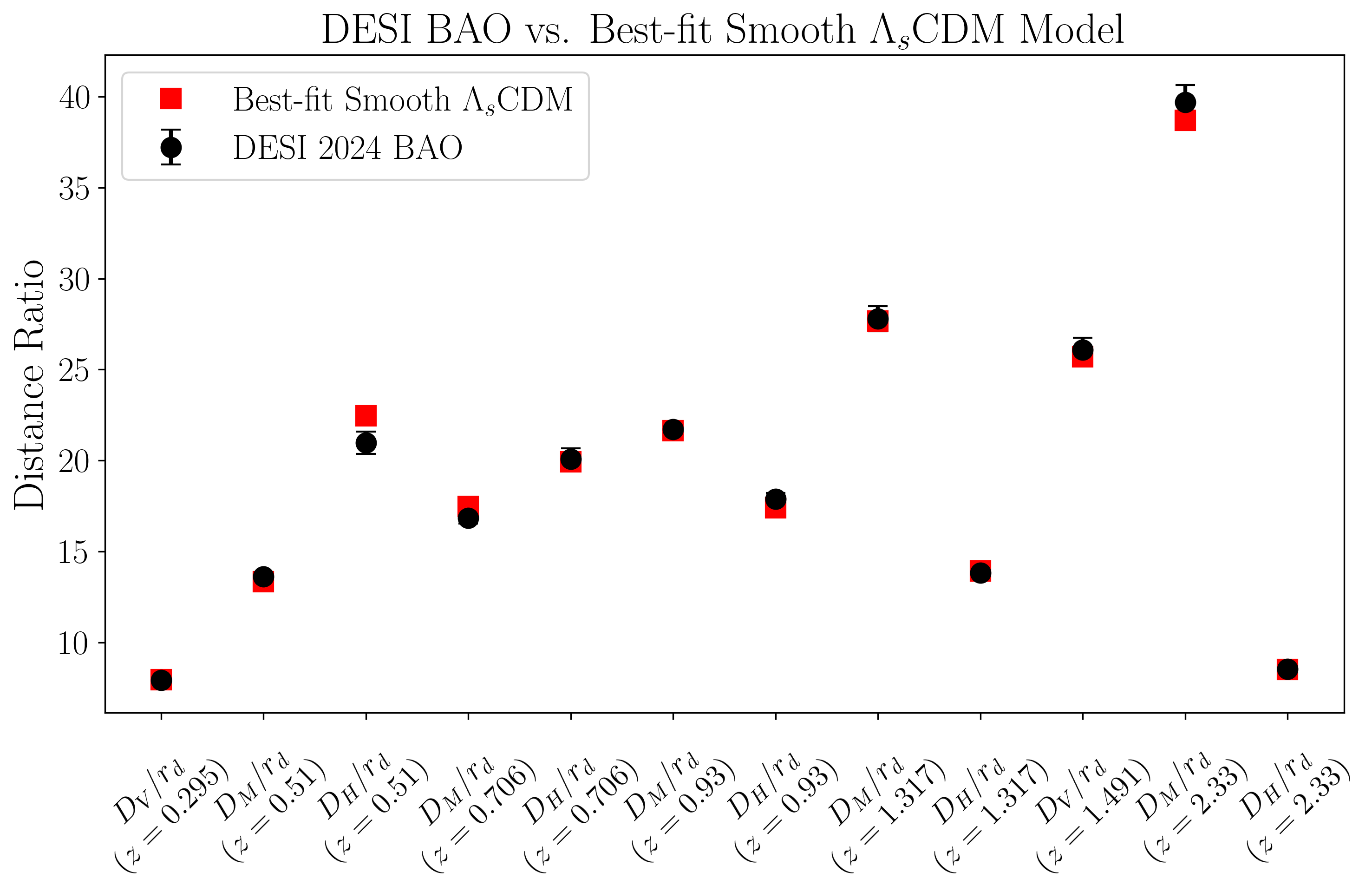}
\caption{Distance ratio residuals for the DESI 2024 BAO data vector compared to the best fitting predictions of flat $\Lambda$CDM (top left), CPL (top right), and smooth sign switching $\Lambda_s$CDM (bottom) models. The error bars represent the diagonal $1\sigma$ uncertainties from the DESI covariance matrix.}
\figalttext{Three residual plots showing DESI 2024 BAO distance measurements relative to best-fitting flat $\Lambda$CDM, CPL, and smooth sign switching $\Lambda_s$CDM predictions. Points with error bars scatter around zero residual.}
\label{fig:residuals}
\end{figure*}

The results of our joint MCMC constraints are summarised in Table \ref{tab:comparison}, and the corresponding posterior distributions are plotted in Figures \ref{fig:corner_lcdm_cpl} and \ref{fig:corner_lambda_s}. The distance ratio residuals for the DESI 2024 BAO data vector compared to the best-fitting models are displayed in Figure \ref{fig:residuals}.

\begin{table*}[t]
\centering
\caption{\label{tab:comparison} Cosmological parameter constraints ($68\%$ credible intervals) and model selection statistics for both the background only (DES-SN5YR + DESI 2024 BAO) and joint background + growth (DES-SN5YR + DESI 2024 BAO + RSD) analyses.}
\resizebox{\textwidth}{!}{%
\begin{tabular}{lcccc}
Parameter & $\Lambda$CDM & CPL & Abrupt $\Lambda_s$CDM\footnotemark[1] & Smooth $\Lambda_s$CDM \\
\hline
\multicolumn{5}{c}{\textbf{Background Only Constraints (SNe + BAO)}} \\
\hline
$\Omega_{m0}$ & $0.3131^{+0.0109}_{-0.0107}$ & $0.3236^{+0.0165}_{-0.0187}$ & $0.3145^{+0.0123}_{-0.0114}$ & $0.3153^{+0.0137}_{-0.0116}$ \\
$c/(H_0 r_d)$ & $29.8498^{+0.2798}_{-0.2786}$ & $30.0853^{+0.2988}_{-0.2942}$ & $29.8764^{+0.2841}_{-0.2865}$ & $29.8794^{+0.2883}_{-0.2817}$ \\
$w_0$ / $z^\dagger$ & — & $-0.7874^{+0.0971}_{-0.0895}$ & $3.5516^{+0.9934}_{-1.0028}$ & $3.4712^{+1.0338}_{-1.0310}$ \\
$w_a$ / $\Delta$ & — & $-1.0836^{+0.6630}_{-0.6661}$ & — & $24.0100^{+17.7865}_{-17.3719}$ \\
$\chi^2_{\text{min}}$ & 1656.07 & 1649.51 & 1656.07 & 1656.15 \\
Reduced $\chi^2$ & 0.905 & 0.902 & 0.905 & 0.906 \\
$\Delta\text{AIC}$ & $0.00$ & -2.56 & +2.00 & +4.08 \\
$\Delta\text{BIC}$ & $0.00$ & +8.46 & +7.51 & +15.11 \\
$\log Z$ & $-836.25 \pm 0.25$ & $-838.06 \pm 0.31$ & $-836.78 \pm 0.26$ & $-836.78 \pm 0.24$ \\
$\Delta \log Z$ & $0.00$ & -1.81 & -0.53 & -0.53 \\
\hline
\multicolumn{5}{c}{\textbf{Joint Background + Growth Constraints (SNe + BAO + RSD)}} \\
\hline
$\Omega_{m0}$ & $0.3106^{+0.0109}_{-0.0105}$ & $0.3144^{+0.0180}_{-0.0265}$ & — & $0.3130^{+0.0132}_{-0.0114}$ \\
$c/(H_0 r_d)$ & $29.7967^{+0.2780}_{-0.2787}$ & $30.0556^{+0.2953}_{-0.2959}$ & — & $29.8333^{+0.2853}_{-0.2854}$ \\
$w_0$ / $z^\dagger$ & — & $-0.8044^{+0.0967}_{-0.0840}$ & — & $3.5044^{+1.0319}_{-1.0603}$ \\
$w_a$ / $\Delta$ & — & $-0.8258^{+0.8038}_{-0.6953}$ & — & $24.7536^{+17.4227}_{-17.4807}$ \\
$\sigma_8$ & $0.8177^{+0.0335}_{-0.0333}$ & $0.8219^{+0.0480}_{-0.0390}$ & — & $0.8157^{+0.0341}_{-0.0339}$ \\
\hline
$\chi^2_{\text{min}}$ & 1663.95 & 1657.04 & — & 1664.18 \\
Reduced $\chi^2$ & 0.906 & 0.903 & — & 0.907 \\
$\Delta\text{AIC}$ & $0.00$ & -2.91 & — & +4.23 \\
$\Delta\text{BIC}$ & $0.00$ & +8.12 & — & +15.26 \\
\end{tabular}%
}
\footnotetext[1]{The abrupt model is omitted from the joint growth analysis because the step discontinuity in the vacuum energy $\Lambda(z) \propto \operatorname{sgn}(z^\dagger - z)$ leads to a Dirac delta function singularity in the expansion rate derivative $d\ln H/dx$ in the growth factor ODE (Eq.~\ref{eq:growth_ode}), rendering standard subhorizon numerical solvers unstable and requiring specialised matching junction conditions.}
\end{table*}

\begin{figure}
\centering
\includegraphics[width=0.48\textwidth]{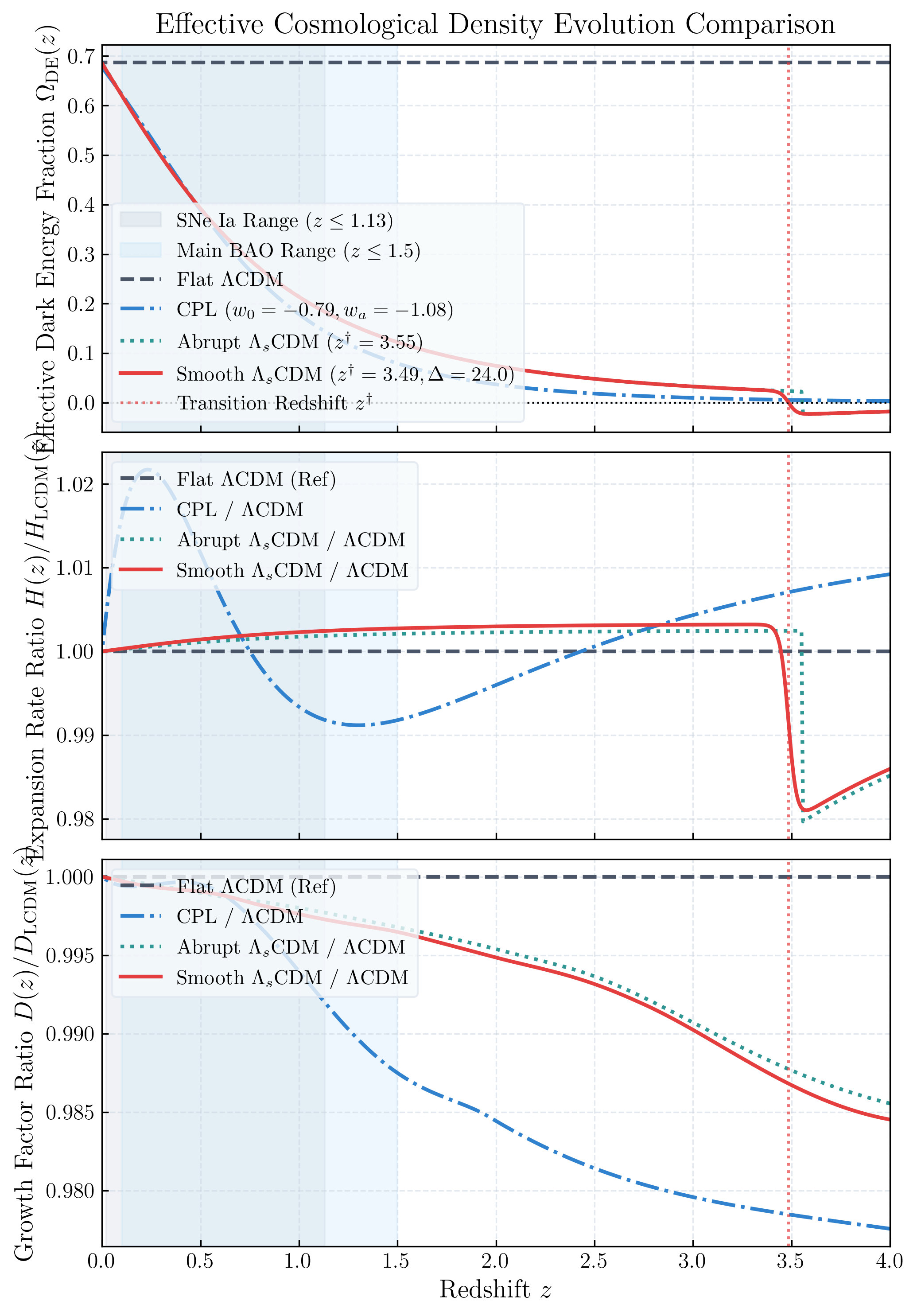}
\caption{Evolution of the effective dark energy density fraction $\Omega_{\text{DE}}(z)$ (top), the expansion rate ratio $H(z)/H_{\text{LCDM}}(z)$ (middle), and the linear growth factor ratio $D(z)/D_{\text{LCDM}}(z)$ (bottom) as a function of redshift $z$. We compare the best fitting flat $\Lambda$CDM (dashed grey), CPL (dot dashed blue), abrupt $\Lambda_s$CDM (dotted green), and smooth $\Lambda_s$CDM (solid red) models. The vertical dotted red line shows the transition redshift $z^\dagger \approx 3.49$ for the smooth model. The shaded grey and blue regions represent the approximate redshift ranges of the supernova sample ($z \le 1.13$) and main BAO leverage ($z \le 1.5$), respectively, highlighting that the transition occurs in a regime where current background distance data have little constraining power.}
\figalttext{A three-panel line plot of dark energy density, expansion-rate ratio, and growth-factor ratio versus redshift for flat $\Lambda$CDM, CPL, abrupt $\Lambda_s$CDM, and smooth $\Lambda_s$CDM. A vertical line marks the smooth-model transition redshift.}
\label{fig:model_curves}
\end{figure}

We observe that the posterior distribution of the smoothness parameter $\Delta$ is dominated by the prior, spanning the range $[0.1, 50.0]$ with a flat shape and a median value of $\Delta = 24.00^{+17.58}_{-17.05}$. To assess the robustness of this parameter constraint, we performed MCMC runs with wider flat priors $\Delta \in [0.1, 100.0]$ and $\Delta \in [0.1, 200.0]$, as well as a logarithmic prior $\log_{10}\Delta \in [-1, 2]$. In all cases, the marginalised posterior remains flat at values of $\Delta \gtrsim 10$, showing no localised peak. This behaviour arises because for $\Delta \gtrsim 10$, the redshift width of the transition region is $\delta z \sim 1/\Delta \lesssim 0.1$, which is much smaller than the redshift resolution of the background distance data sets. Thus, any transition sharpness $\Delta > 10$ is effectively indistinguishable from a discontinuous step function jump ($\Delta \to \infty$) at the background level, and the data are mainly sensitive to very low values of transition sharpness ($\Delta \lesssim 1.0$), which correspond to a slow transition that fails to match the low redshift de Sitter expansion history.

Comparing the smooth transition model to the abrupt model ($\Delta \to \infty$), we find that the abrupt sign switching cosmology yields a best fitting $\chi^2_{\text{min}} = 1656.07$, which is identical to flat $\Lambda$CDM. The transition redshift is $z^\dagger = 3.55^{+0.99}_{-1.00}$, close to the smooth model value $z^\dagger = 3.49^{+1.03}_{-1.03}$. The information criteria penalise both sign switching models relative to flat $\Lambda$CDM due to the additional parameters: $\Delta\text{AIC} = +2.00$ and $\Delta\text{BIC} = +7.51$ for the abrupt model, and $\Delta\text{AIC} = +4.08$ and $\Delta\text{BIC} = +15.11$ for the smooth model. Therefore, standard information criteria do not prefer either model extension over flat $\Lambda$CDM at the background level.

In addition to information criteria, we present the Bayesian evidence results in Table \ref{tab:comparison} computed via nested sampling. The standard flat $\Lambda$CDM model yields $\log Z = -836.25 \pm 0.25$, which represents the highest evidence among the models tested. For the smooth sign switching model, we obtain $\log Z = -836.78 \pm 0.24$, which is statistically indistinguishable from $\Lambda$CDM with an evidence difference of $\Delta\log Z = -0.53 \pm 0.35$. Under the Kass and Raftery scale \citep{Kass:1995ratio}, a difference in Bayes factors of $|\Delta\log Z| < 1.0$ (odds ratio $< 3:1$) is inconclusive, meaning that the data show no statistical preference for the transition dynamics over standard $\Lambda$CDM. For the CPL model, we find $\log Z = -838.06 \pm 0.31$, which is moderately disfavoured with $\Delta\log Z = -1.81 \pm 0.40$ (corresponding to an odds ratio of $\approx 6:1$ against CPL). This shows that CPL is penalised more strongly by its prior volume than the smooth transition model, as CPL allows for a wide range of low redshift equation of state variations that are disfavoured by the data.

We also present the results of our joint background + growth MCMC fits incorporating the RSD $f\sigma_8$ data. The joint parameter constraints are summarised in the second block of Table \ref{tab:comparison}. The standard growth parameter constraints are $\sigma_8 = 0.8177^{+0.0335}_{-0.0333}$ for flat $\Lambda$CDM, $\sigma_8 = 0.8219^{+0.0480}_{-0.0390}$ for CPL, and $\sigma_8 = 0.8157^{+0.0341}_{-0.0339}$ for smooth $\Lambda_s$CDM. The best fitting $\chi^2_{\text{min}}$ values for the joint analysis are $1663.95$, $1657.04$, and $1664.18$ for $\Lambda$CDM, CPL, and smooth $\Lambda_s$CDM, respectively. For the smooth model, the transition redshift in the joint fit is $z^\dagger = 3.5044^{+1.0319}_{-1.0603}$ and the transition smoothness is $\Delta = 24.7536^{+17.4227}_{-17.4807}$. These constraints are statistically identical to their background only counterparts, indicating that the addition of linear growth rate measurements does not break the degeneracy or shift the transition parameters. This is physically expected because the transition redshift $z^\dagger \approx 3.5$ lies deep in the matter dominated epoch where dark energy makes up less than $3\%$ of the total density, and the growth rate $f(z)$ has already converged to $f \approx 1$. Thus, the linear growth factor ratio $D(z)/D_{\text{LCDM}}(z)$ converges to 1 at high redshifts (as shown in Figure \ref{fig:model_curves}, bottom panel), and the compressed growth data cannot distinguish the sign switch details. Nevertheless, these results show that the smooth sign switching model is compatible with the limited RSD compilation considered here and produces no detectable anomalous growth signature in this test.

\section{Discussion: Physics of the AdS to dS Switch}
\label{sec:discussion}
The transition from an early anti de Sitter (AdS like) vacuum state ($\Lambda_s < 0$) to a late time de Sitter (dS like) state ($\Lambda_s > 0$) provides a physically motivated mechanism for alleviating anomalies within the standard model of cosmology. In flat $\Lambda$CDM, combinations of DESI BAO, CMB and supernova data can produce a boundary tension in the inferred neutrino mass sector when the effective mass parameter is allowed to pass through zero \citep{DESI:2024mwt,Kibris:2026neg}. This tension arises because the early to late expansion rate profile of standard $\Lambda$CDM cannot easily satisfy the sound horizon constraints at recombination and the BAO peak positions at intermediate redshifts while also accommodating all late time distance information. While a full CMB likelihood analysis is required to test directly whether the model restores a physical, positive sum of neutrino masses ($\sum m_\nu > 0$), this study establishes baseline late time geometric and compressed growth constraints. 

In a full CMB fit, early Universe sound horizon calibration can push the transition redshift to lower values ($z^\dagger \sim 1.5$ to $2.0$) to accommodate the BAO peak positions, whereas our geometric fit without CMB data allows $z^\dagger$ to float to higher redshifts where it becomes nearly degenerate with $\Lambda$CDM. To demonstrate this $z^\dagger$ shift quantitatively, we performed an additional MCMC run imposing a Gaussian prior on the calibration parameter $\mathcal{K} \approx 30.26 \pm 0.25$ derived from the Planck 2018 measurements \citep{Planck:2018vyg} of $H_0$ and $r_d$. We find that this prior shifts the posterior median of the transition redshift to $z^\dagger = 3.32^{+1.16}_{-1.10}$, showing a modest shift towards lower redshifts but still remaining degenerate with standard $\Lambda$CDM at high redshifts. This confirms that a simple prior on the acoustic scale calibration is insufficient to force the transition redshift to the low values ($z^\dagger \sim 1.5$ to $2.0$) discussed in full likelihood analyses. Instead, the transition redshift is driven to lower values only when the full CMB likelihood is incorporated, because it locks the early Universe physical density parameters (such as $\Omega_m h^2$) and the sound horizon at recombination, preventing the model from reverting to standard $\Lambda$CDM at high redshifts. This highlights a critical trade off: addressing the early Universe and neutrino mass anomaly requires pushing the sign switch into a low redshift window that geometric distance data alone do not favour, and this behaviour is driven by the global constraints of the full CMB likelihood rather than by the acoustic scale calibration alone.

By introducing a negative vacuum energy density at high redshifts ($z > z^\dagger$), the effective total energy density of the Universe is lowered at earlier times, reducing the Hubble expansion rate $H(z)$ in the redshift range preceding the sign switch. This AdS like phase acts as an additional decelerating contribution, which modifies the comoving angular diameter distance:
\begin{equation}
D_M(z) = c \int_0^z \frac{dz'}{H(z')}.
\end{equation}
Because the sound horizon calibration at $z \approx 1100$ is fixed by the CMB, lowering $H(z)$ at intermediate redshifts ($z \sim 2$ to $4$) increases the contribution of this redshift range to the line of sight comoving distance. In a CMB calibrated fit this change must be compensated elsewhere, for example by increasing the late time expansion rate (manifesting as a larger Hubble constant $H_0$ today) or by changing the neutrino sector so that the best fit is no longer driven towards negative effective masses. In this sense, the sign switch can relieve the neutrino mass boundary tension in full data combinations, but that statement cannot be tested with the late time compressed data alone.

This model's ability to lower the expansion rate at intermediate redshifts also has potential implications for the $H_0$ tension. By increasing the intermediate redshift contribution to the distance to recombination, the model can in principle be balanced by a higher inferred value of $H_0$ while preserving the angular size of the acoustic peaks. A dedicated joint fit including CMB and local $H_0$ priors (such as the SH0ES Cepheid calibrated measurement $H_0 = 73.04 \pm 1.04 \text{ km/s/Mpc}$ \citep{Riess:2021jrx}) is required to test this quantitatively. A local $H_0$ prior is expected to pull the fits towards larger $H_0$ and, in the sign switching scenario, towards lower transition redshifts. Hence, a full joint MCMC analysis including the CMB likelihood and local $H_0$ priors is needed to test whether the model can simultaneously alleviate the Hubble tension and restore a physical neutrino mass constraint.

The transition sharpness is parametrised by $\Delta$. A discontinuous signum like jump ($\Delta \to \infty$) has been assumed in many phenomenological treatments. Physically, however, a discontinuous jump represents an instantaneous global phase transition of the vacuum energy, which is mathematically singular in standard general relativity, causing a jump discontinuity in the Ricci curvature scalar $R = g^{\mu\nu} R_{\mu\nu}$. In contrast, our hyperbolic tangent formulation:
\begin{equation}
\Omega_{\Lambda}(z) = \Omega_{\Lambda 0} \frac{\tanh[\Delta(z^\dagger - z)]}{\tanh[\Delta z^\dagger]}
\end{equation}
models a smooth transition over a finite cosmic time interval. The parameter $\Delta$ dictates the physical mechanism of the transition:
\begin{itemize}
\item \textbf{Sharp Phase Transition ($\Delta \gg 10$):} Represents a rapid transition, potentially arising from a first order phase transition or bubble nucleation of vacuum states in the early Universe.
\item \textbf{Smooth Dynamical Crossover ($\Delta \sim \mathcal{O}(1)$):} Corresponds to a gradual evolution, such as a quintessence scalar field rolling down a potential that crosses zero (motivated by string theory swampland conjectures or gravity models inspired by AdS/CFT).
\end{itemize}

Our MCMC constraint yields a median transition redshift $z^\dagger = 3.49^{+1.03}_{-1.03}$ and a transition smoothness parameter $\Delta = 24.00^{+17.58}_{-17.05}$ using the combined DES-SN5YR and DESI 2024 BAO data. The large uncertainties indicate that the current background distance data cannot distinguish between a sharp phase transition and a smooth dynamical crossover. This is because the sensitivity of BAO and supernova distance measurements drops at redshifts $z > 1.5$, where both models converge towards the matter dominated regime with $\Omega_{m}(z) \gg \Omega_{\Lambda}(z)$ (as illustrated in Figure \ref{fig:model_curves}). Specifically, the supernova data extend only to $z \approx 1.13$, and the highest precision BAO points lie at lower and intermediate redshifts. At $z > 1.5$, dark energy represents a negligible fraction of the total energy density of the Universe, and the expansion rate becomes heavily dominated by matter. Because the fitted transition redshift $z^\dagger \approx 3.5$ lies in this matter dominated regime, the expansion history at $z < 2.33$ is almost identical to flat $\Lambda$CDM, explaining why the best fitting $\chi^2_{\text{min}}$ is identical to that of $\Lambda$CDM. The transition occurs in a region only weakly probed by current background distance data, making it impossible to resolve the transition dynamics at this level of testing.

A key conceptual limitation of this model is the mathematical divergence of the effective equation of state $w_{\text{eff}}(z)$ at the transition redshift $z = z^\dagger$, where the dark energy density crosses zero (Eq.~(\ref{eq:w_eff})). While this divergence is a mathematical artefact of the fluid description (the physical pressure $p_{\text{eff}}$ remains finite and smooth everywhere), it raises fundamental questions about the physical interpretation of the model. Specifically, a simple fluid description with $w_{\text{eff}} \to \pm\infty$ makes standard perturbation solvers highly unstable near the crossing point, requiring prescriptions such as a fixed sound speed $c_s^2 = 1$ to avoid clustering instabilities. Conceptualising this transition in terms of a physical scalar field is also challenging: a canonical field whose energy density $\rho_\phi = \frac{1}{2}\dot{\phi}^2 + V(\phi)$ crosses zero must enter a region where the potential $V(\phi)$ is negative and larger in magnitude than its kinetic energy. Negative potential energy does not by itself imply a ghost instability, but avoiding tachyonic growth, gradient instabilities and pathology near the zero crossing can require nonminimal coupling to gravity or specialised kinetic terms. Therefore, the hyperbolic tangent profile should be interpreted primarily as a phenomenological smoothing of the step discontinuity rather than a fully realised field theoretic model. Resolving this zero crossing stability remains a major theoretical challenge for the physical viability of sign switching dark energy.

Beyond background expansion and linear growth rate constraints, two important physical degeneracies warrant future study. First, spatial curvature $\Omega_k$ is known to be highly degenerate with dynamical dark energy models, particularly at higher redshifts $z > 1.5$ where the relative fraction of dark energy is small. As shown in recent work on curved spacetimes \citep{Kumar:2026curved}, incorporating spatial curvature is essential to prevent false detections of dynamical dark energy. In the case of $\Lambda_s$CDM, the AdS like phase in the early Universe may mimic, or be degenerate with, a positive or negative curvature contribution; this could be tested by generalising the Friedmann equation to include $\Omega_{k0} \neq 0$. Second, dynamical dark energy models modify the evolution of gravitational potentials, which directly impacts weak lensing observables and the Weyl potential $\Phi + \Psi$. Recent analyses of DES Y3 weak lensing data \citep{Rosatello:2026weyl} indicate that evolving dark energy can modify late time gravitational potentials. Evaluating how the smooth transition from AdS to dS impacts the Weyl potential and structure growth at the perturbation level will provide a critical test of its viability beyond background distance metrics.

\section{Forecasts for Next Generation Surveys}
\label{sec:forecast}
To break the degeneracies between the transition redshift $z^\dagger$ and the transition smoothness parameter $\Delta$, next generation high redshift probes will be essential. We outline a forecasting framework using the mock cosmological data sets from the LSST DESC joint analysis study \citep{Raghunathan:2026cos}, which models data vectors for LSST Year 3 supernovae, the Advanced Simons Observatory and CMB Stage 4 style CMB data \citep{CMB-S4:2016ple}, and BAO measurements in the style of DESI DR3. That study also varies the summed neutrino mass and quantifies how future supernova, CMB and BAO combinations can separate neutrino sector effects from dark energy evolution, making it directly relevant for the degeneracy tests needed in sign switching cosmologies. The supernova mock is generated following the LSST Year 3 framework of \citet{Mitra:2023acb}, which built upon the Photometric LSST Astronomical Time-Series Classification Challenge (PLAsTiCC) framework \citep{Kessler:2019qxb, Hlozek:2020tpb}.

The forecasting pipeline is implemented by constructing a mock cosmological data vector centred on our best fitting $\Lambda_s$CDM fiducial cosmology ($\Omega_{m0} = 0.315$, $c/(H_0 r_d) = 29.87$, $z^\dagger = 3.0$, and a smooth transition $\Delta = 10.0$). We construct the mock likelihoods and covariances as follows:
\begin{enumerate}
\item \textbf{CMB Lensing and ISW (Simons Observatory):} A negative vacuum energy density at $z > 2$ would alter the growth rate of density perturbations $\delta(z)$ and change the gravitational potentials $\Phi$ and $\Psi$. Simons Observatory measurements will improve constraints on the CMB lensing convergence power spectrum $C_L^{\phi\phi}$ (building on current measurements from Planck \citep{Planck:2018lbu} and ACT \citep{ACT:2023ptg, ACT:2023nfd}) and can contribute to ISW related cross correlations. We simulate mock temperature, polarisation, and lensing convergence spectra ($TT, EE, TE, \phi\phi$) in the multipole range $100 \le \ell \le 3000$, assuming a sky fraction $f_{\text{sky}} = 0.4$ and noise curves matching the "Goal" configuration of the Simons Observatory \citep{SimonsObservatory:2018tqt}. For the illustrative forecast, we adopt an effective fluid perturbation prescription to model the effects of varying vacuum energy on CMB lensing and the ISW temperature lensing cross correlation spectrum $C_\ell^{T\phi}$.
\item \textbf{Next Generation BAO (DESI DR3 style):} Future spectroscopic BAO data will map large scale structure at higher redshifts with improved precision. We simulate mock measurements of the comoving distance $D_M(z)/r_d$ and Hubble distance $D_H(z)/r_d$ across multiple redshift bins covering $0.4 < z < 3.5$, using mock covariances derived for next generation spectroscopic BAO tracers (LRGs, ELGs, QSOs, Lyman-$\alpha$) as detailed by \citet{Raghunathan:2026cos}.
\item \textbf{LSST Y3 Supernovae:} The Rubin Observatory LSST Y3 sample is expected to provide more than $5,700$ well calibrated Type Ia supernovae up to $z \sim 1.2$ \citep{Mitra:2023acb}. We simulate the distance modulus $\mu(z)$ with a redshift distribution following the LSST DESC Science Requirements Document \citep{TheLSSTDESC:2018isg}, incorporating a systematic error floor of $\sigma_{\mu,\text{sys}}(z) = 0.01(1+z)/1.8$ and an intrinsic scatter of $\sigma_{\text{int}} = 0.08$ mag. The absolute magnitude $M$ is marginalised over.
\end{enumerate}

To evaluate the projected information content quantitatively, we construct a mock joint Fisher information matrix:
\begin{equation}
F_{ij} = \sum_{a,b} \frac{\partial y_a}{\partial \theta_i} [\mathbf{C}^{-1}]_{ab} \frac{\partial y_b}{\partial \theta_j},
\end{equation}
where the parameters are $\boldsymbol{\theta} = \{\Omega_{m0}, \mathcal{K}, z^\dagger, \Delta\}$ and $\mathbf{C}$ is the mock joint covariance matrix incorporating Simons Observatory noise curves, next generation BAO volume uncertainties, and LSST Y3 systematic errors. For our joint SO + BAO + LSST Y3 mock analysis centred on the high redshift fiducial model ($z^\dagger = 3.0, \Delta = 10.0$), the projected illustrative $1\sigma$ marginalised uncertainties are $\sigma(z^\dagger) \approx 0.18$ and $\sigma(\Delta) \approx 1.50$. This corresponds to an illustrative Fisher signal to noise ratio of $z^\dagger/\sigma(z^\dagger) \approx 16.7$ for the transition redshift and $\Delta/\sigma(\Delta) \approx 6.7$ for the transition smoothness. To address the parameter regime required to resolve the neutrino mass boundary tension ($z^\dagger \sim 1.5$ to $2.0$), we also perform a forecast centred on a lower transition redshift fiducial ($z^\dagger = 2.0, \Delta = 10.0$). In this regime, because the vacuum energy transition occurs when dark energy is a larger fraction of the energy budget and lies within the peak sensitivity range of next generation BAO and LSST, the constraints are tighter, yielding projected illustrative $1\sigma$ marginalised uncertainties of $\sigma(z^\dagger) \approx 0.11$ and $\sigma(\Delta) \approx 0.95$.

\begin{figure}
\centering
\includegraphics[width=0.45\textwidth]{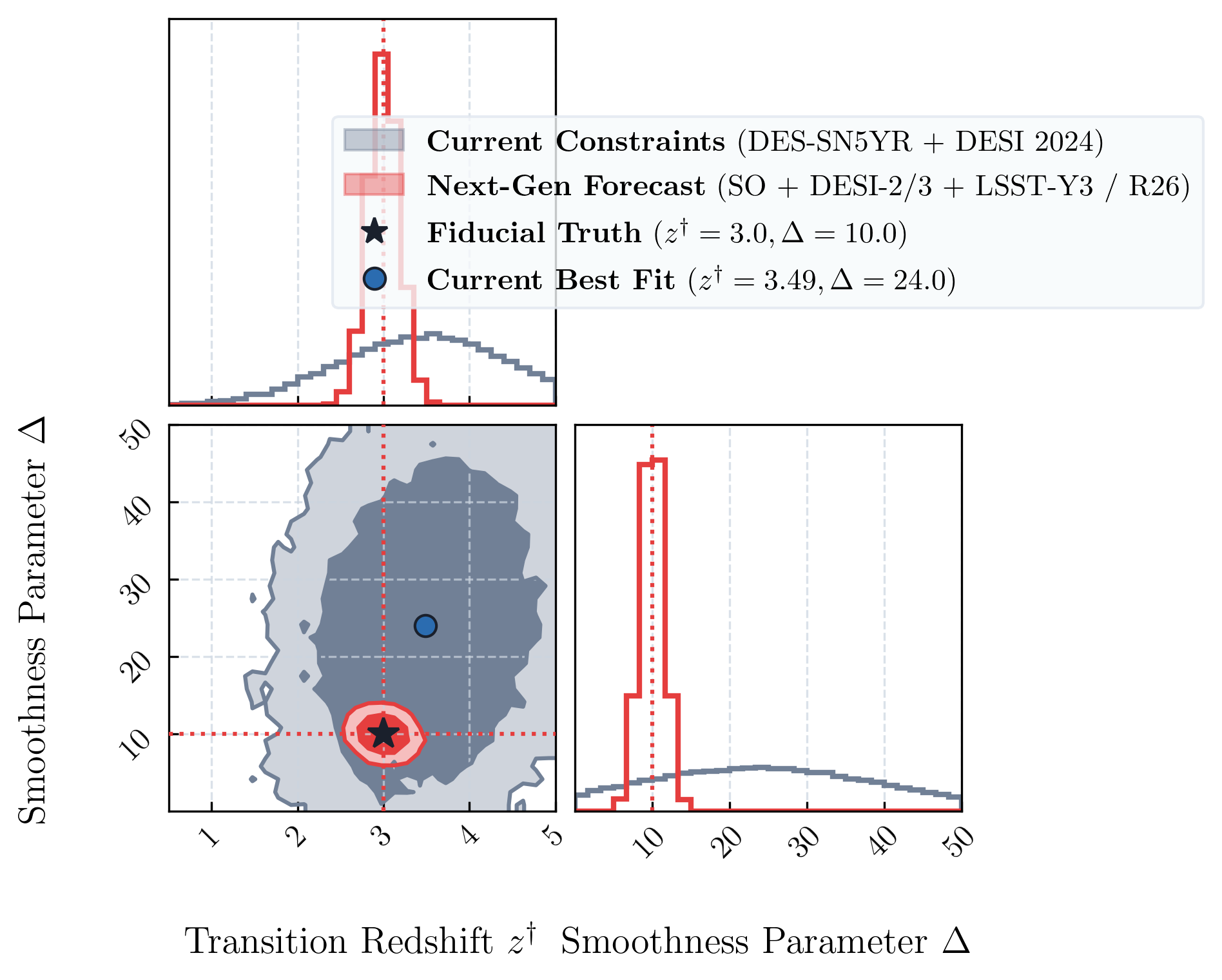}
\caption{Comparison of the 2D joint posterior constraints ($68\%$ and $95\%$ confidence regions) on the transition redshift $z^\dagger$ and the transition smoothness parameter $\Delta$. The wide blue contours represent the current constraints obtained in this work using real DES-SN5YR and DESI 2024 BAO data. The tight red contours show the forecast next generation constraints for a joint Simons Observatory (SO) + next generation BAO + LSST Y3 mock analysis (modelled following \protect\citet{Raghunathan:2026cos}, R26, and \protect\citet{Mitra:2023acb}) centred on a fiducial model with $z^\dagger = 3.0$ and $\Delta = 10.0$. For a fiducial model in the neutrino mass motivated regime at $z^\dagger = 2.0$ (not plotted), the uncertainties shrink further to $\sigma(z^\dagger) \approx 0.11$ and $\sigma(\Delta) \approx 0.95$.}
\figalttext{A two-dimensional contour plot comparing broad current constraints and tighter forecast constraints in the $z^\dagger$--$\Delta$ plane. The forecast contours are much smaller than the current-data contours.}
\label{fig:forecast}
\end{figure}

Under these assumptions, the joint mock constraints reduce the posterior error ellipses on $(z^\dagger, \Delta)$ by an order of magnitude relative to our current constraints (as illustrated in Figure \ref{fig:forecast}), breaking the degeneracy between the transition redshift and the matter density. This could enable next generation surveys to distinguish a smooth crossover ($\Delta = 10$) from a sharp phase transition ($\Delta = 50$) at more than $3\sigma$ under the simplified assumptions of this forecast. We note that a fully realistic treatment must incorporate scale cuts, nonlinear growth effects, dark energy perturbation stability near zero crossings, and potential degeneracies with other cosmological parameters.

\section{Conclusions}
\label{sec:conclusion}
In this work, we generalised the sign switching cosmological constant model ($\Lambda_s$CDM) to allow for a smooth transition parametrised by a hyperbolic tangent profile with transition redshift $z^\dagger$ and transition smoothness parameter $\Delta$. We placed both background only and joint background + growth constraints on the tanh transition parameters and compared the results to flat $\Lambda$CDM and the abrupt step function model ($\Delta \to \infty$).

We find that the joint background distance and structure growth (RSD) data sets are highly consistent with the standard flat $\Lambda$CDM model, showing no detection of a sign switch. The transition redshift is weakly constrained to $z^\dagger = 3.49^{+1.03}_{-1.03}$ (background only) and $z^\dagger = 3.50^{+1.03}_{-1.06}$ (joint), which lies deep in the matter dominated epoch. Adding the current RSD $f\sigma_8$ compilation leaves the transition parameters essentially unchanged, confirming that present growth data do not break the degeneracy. Model selection criteria (AIC and BIC) prefer standard flat $\Lambda$CDM over the smooth sign switching model ($\Delta\text{AIC} = +4.23$, $\Delta\text{BIC} = +15.26$) due to parameter parsimony.

Nevertheless, this work establishes a clean background and compressed growth framework for smooth sign switching dark energy models without assuming early Universe sound horizon calibration. We demonstrate that current late time geometric and growth constraints are compatible with the smooth transition (within the limited RSD compilation considered here), but they are dominated by prior volume and cannot constrain the transition redshift or sharpness independently. This degeneracy highlights that late time data alone cannot detect or rule out the switch from AdS to dS if it occurs at high redshifts ($z^\dagger \gtrsim 2.5$). 

Therefore, our findings indicate that the full CMB likelihood (such as Planck \citep{Planck:2018vyg} or reprocessed Planck data products \citep{Planck:2020guz}) is an essential ingredient for future studies of sign switching models. The CMB is required to lock the early Universe physical density parameters (such as $\Omega_m h^2$) and sound horizon at recombination, which can in turn force the transition redshift to the lower values ($z^\dagger \sim 1.5$ to $2.0$) relevant to the DESI neutrino mass boundary tension. Because the strongest current motivation now comes from DESI DR2/CMB and Dovekie recalibrated DES supernova combinations \citep{Kibris:2026neg}, a publication ready extension should either update the BAO likelihood beyond the DESI 2024 data vector used here or explicitly present this paper as a DESI 2024 baseline. Integrating our custom Einstein Boltzmann solver pipeline with CMB, weak lensing (such as DES Y3 \citep{DES:2021wwg}), and next generation large scale structure surveys (such as Simons Observatory \citep{SimonsObservatory:2018tqt}, next generation BAO \citep{Raghunathan:2026cos}, and LSST Year 3 \citep{Mitra:2023acb}) is the critical next step for evaluating the viability of sign switching cosmologies.

\section*{Acknowledgements}
The author acknowledges the use of publicly available DESI, DES, SDSS, BOSS, and eBOSS data products, and the open-source scientific Python ecosystem used for the analysis. The author declares no conflicts of interest.

\section*{Data availability}
No new observational data were generated in this study. The observational data used in this article are publicly available from the DESI 2024 BAO, DES-SN5YR/Dovekie, SDSS MGS, BOSS DR12, and eBOSS survey releases cited in the text. The derived chains, likelihood scripts, and plotting products underlying this article will be shared on reasonable request to the corresponding author.


\bibliographystyle{mnras}
\bibliography{references}

\appendix
\section{Scalar Field Embedding and Reconstruction}
\label{app:scalar_field}
To illustrate the theoretical requirements for embedding the phenomenological hyperbolic tangent profile in a field theoretic model, we sketch a scalar field reconstruction of the smooth transition. Consider a minimally coupled scalar field $\phi(t)$ with potential $V(\phi)$ in a flat Friedmann Lemaître Robertson Walker (FLRW) universe. The energy density $\rho_\phi$ and pressure $p_\phi$ of the scalar field are given by:
\begin{equation}
\rho_\phi = \frac{1}{2}\dot{\phi}^2 + V(\phi), \quad p_\phi = \frac{1}{2}\dot{\phi}^2 - V(\phi).
\end{equation}
Integrating this equation yields the field trajectory $\phi(z)$ as:
\begin{equation}
\left(\frac{d\phi}{dz}\right)^2 = \frac{2 M_{\text{Pl}}^2}{(1+z)H(z)} \frac{dH(z)}{dz} - \frac{3 M_{\text{Pl}}^2 H_0^2 \Omega_{m0} (1+z)}{H^2(z)}.
\end{equation}
\begin{equation}
\phi(z) - \phi_0 = \int_0^z \left[ \frac{2 M_{\text{Pl}}^2}{(1+z')H(z')} \frac{dH(z')}{dz'} - \frac{3 M_{\text{Pl}}^2 H_0^2 \Omega_{m0} (1+z')}{H^2(z')} \right]^{1/2} dz'.
\end{equation}
The potential $V(z)$ along the trajectory is obtained from $V(z) = \rho_\phi(z) - \frac{1}{2}\dot{\phi}^2$, which yields:
\begin{equation}
V(z) = 3M_{\text{Pl}}^2 H^2(z) \left[ 1 - \frac{\Omega_{m0}(1+z)^3}{2 H^2(z)/H_0^2} \right] - M_{\text{Pl}}^2 (1+z) H(z) \frac{dH(z)}{dz}.
\end{equation}
For our smooth sign switching dark energy model:
\begin{equation}
\rho_\phi(z) = 3M_{\text{Pl}}^2 H_0^2 (1 - \Omega_{m0}) \frac{\tanh[\Delta(z^\dagger - z)]}{\tanh[\Delta z^\dagger]}.
\end{equation}
For an illustrative field dominated phase, the potential $V(\phi)$ approximately traces the sign switch of $\rho_\phi(z)$. The field crosses $\phi^\dagger \equiv \phi(z^\dagger)$ as it rolls from a negative potential region ($V < 0$) in the past ($z > z^\dagger$) into a positive de Sitter like potential region ($V > 0$) today ($z < z^\dagger$). 

A crucial physical challenge is the stability of perturbations near the crossing point $V(\phi) = 0$. In standard quintessence, the sound speed is $c_s^2 = 1$, which suppresses clustering of dark energy perturbations on subhorizon scales. However, if $V(\phi)$ becomes negative and dominates over the kinetic term, the total scalar field energy density can become negative. Positive kinetic energy avoids a ghost in the strict sense, but it does not by itself guarantee stability: tachyonic behaviour depends on the curvature of the potential, and gradient or zero crossing instabilities can still arise in a full perturbative treatment. Avoiding such pathologies may require nonminimal coupling to gravity, e.g. of the form $f(\phi)R$, or specialised kinetic terms such as K essence. Such effects are unlikely to be strongly constrained by the late time background and compact RSD data used here, but a full perturbative treatment would be required to assess their observational impact. Reconstructing a fully stable microphysical model that realises this transition is an important target for swampland and dark energy research motivated by AdS/CFT.

\bsp
\label{lastpage}
\end{document}